\documentclass[12pt]{article}
\usepackage{amssymb}
\usepackage{amsmath}
\newcommand{\email}[1]{\vspace*{5pt} {E-mail: \tt{#1}}}
\newcommand{\hepth}[1]{{\tt hep-th/#1}}
\newcommand{\pd}{\partial}
\newcommand{\grqc}[1]{{\tt gr-qc/#1}}
\newcommand{\arXivid}[1]{{\tt arXiv:#1}}
\newcommand{\astroph}[1]{{\tt astro-ph/#1}}

\newcommand{\mathph}[1]{{\tt math-ph/#1}}

\newcommand{\Fc}{\mathcal{F}}

\begin{document}

\title{\textbf{Localization of Nonlocal Cosmological\\ Models with Quadratic Potentials\\ in the case of Double Roots}}

\author{\textit{Sergey  Yu. Vernov}\\
{}\\
Skobeltsyn Institute of Nuclear Physics, \\
Moscow State University,\\ Leninskie gory, GSP-1, Moscow, 119991, Russia,\\
\small{\email{svernov@theory.sinp.msu.ru}}}

\date{ }

\maketitle

\begin{abstract}
Nonlocal cosmological models with quadratic potentials are considered.
We study the action with an arbitrary analytic function $\Fc(\Box_g)$,
which has both double and simple roots. The formulae for nonlocal
energy--momentum tensor, which correspond to double roots, have been
obtained. The way to find particular solutions for nonlocal Einstein
equations in the case when $\Fc(\Box_g)$ has both simple and double
roots has been proposed. One and the same functions solve the initial
nonlocal Einstein equations and the obtained local Einstein equations.
\end{abstract}


\section{Introduction}

Recently a wide class of nonlocal cosmological models based on the
string field theory (SFT) (for details see
reviews~\cite{review-sft}) and the $p$-adic string theory
\cite{padic} emerges and attracts a lot of attention
\cite{IA1}--\cite{CN}. The SFT inspired cosmological models are
intensively considered as models for  dark energy (DE). Actions of
some of cosmological models originating from the SFT have terms
with infinite order derivatives, i.e. nonlocal terms.

Due to the presence of phantom excitations nonlocal models are of
interest for the present cosmology. The inequality
$w_{\mathrm{DE}}<-1$, where $w_{\mathrm{DE}}$ is the DE state
parameter, means the violation of the null energy condition (NEC).
Field theories which violate the null energy condition are
actively studied as a possible solution of the cosmological
singularity problem \cite{Hawking-Ellis,cyclic,GV} and as models
of  dark energy (see \cite{Caldwell}--\cite{Carroll} and
references therein). Generally speaking, models that violate the
NEC have ghosts, and therefore  are unstable and physically
unacceptable. Phantom fields look harmful to the theory and a
local model with a phantom scalar field is not acceptable from the
general point of view. Models with higher derivative terms produce
well-known problems with quantum instability~\cite{AV-NEC,RAS}.
Several attempts to solve these problems have been recently
performed~\cite{SW,Creminelli0812}.  A  physical idea that could
 solve the problems
is  that  the instabilities do not have  enough time to fully
develop. A mathematical one is that dangerous terms can be treated
as corrections valued only at small energies below the physical
cut-off. This approach implies the possibility  to construct a UV
completion of the theory, and this assumption requires detailed
analysis.

Note that the possibility of the existence of dark energy with
$w_{\mathrm{DE}}<-1$ is not excluded experimentally. Indeed,
contemporary cosmological observational data~\cite{data} strongly
support that the present Universe exhibits an accelerated
expansion providing thereby an evidence for a dominating DE
component (for a review see also~\cite{review-de}). Recent results
of WMAP together with the data on Ia supernovae give the following
bounds for the DE state parameter
\begin{equation}
w_{\mathrm{DE}}={} -1.0\pm 0.2.
\end{equation}
The present cosmological observations do not exclude an evolving
parameter~$w_{\mathrm{DE}}$. Moreover, the recent analysis of the
observation data indicates that the varying in time dark energy with
the state parameter $w_{DE}$, which crosses the cosmological constant
barrier, gives a better fit than a cosmological
constant~\cite{ZhangGui} (for details see reviews~\cite{Quinmodrev1}
and references therein).

To obtain a stable model with $w_{\mathrm{DE}}<-1$ one should construct
the effective theory with the NEC violation from the fundamental
theory, which is stable and admits quantization.   From this point of
view the NEC violation might be a property of a model that approximates
the fundamental theory and describes some particular features of the
fundamental theory. With the lack of quantum gravity, we can just trust
string theory or deal with an effective theory admitting  the UV
completion.
 It can be considered as a hint towards the SFT inspired
cosmological models.

Among cosmological models with $w_{\mathrm{DE}}<-1$, which have been
constructed to be free of instability problem, we can mention the
Lorentz-violating dark energy model~\cite{Rubakov}, the $k$-essence
models~(see \cite{Creminelli0812} and references therein) and the
brane-world models~\cite{Brane}.

For a more general discussion on the string cosmology and coming
out of string theory theoretical explanations of the cosmological
observational data the reader is referred
to~\cite{string-cosmo,0605265,Biswas}. Other models obeying
nonlocality and their cosmological consequences are considered in
\cite{nonlocal}. In the flat space-time nonlocal equations are
actively investigated as well~\cite{PaisU,STinspired,Yang,AJK}.
Note that differential equations of infinite order were begin to
study in the mathematical literature long time
ago~\cite{davis,carmi} (see~\cite{noghosts} as a review).

The purpose of this paper is to study the  string field theory inspired
nonlocal model with a quadratic potential. In this paper we consider a
 general form of linear nonlocal action for the scalar field
keeping the main ingredient, the function $\Fc(\Box_g)$, which in fact
produces the nonlocality in question, almost unrestricted. The only
strong restriction we impose is the analyticity of $\Fc(\Box_g)$. In
previous papers~\cite{Koshelev07,AJV0701,AJV0711,MN,KV} only simple
roots have been considered. In this paper we consider the case of the
function $\Fc(\Box_g)$ with both simple and double roots.

 The possible way to find solutions of the Einstein equations
with a quadratic potential of the nonlocal scalar field, is to
reduce them to a system of Einstein equations describing many
non-interacting free local scalar fields~\cite{Koshelev07,AJV0711}
(see also~\cite{KV}). The masses of all local fields are roots of
an algebraic or transcendental equation, which appears in the
nonlocal model. Some of the obtained local scalar fields are
normal and other of them are phantom ones.

The particular forms of  $\Fc(\Box_g)$ are inspired by the fermionic
SFT and the most well understood process of tachyon condensation.
Namely, starting with a non-supersymmetric configuration the tachyon of
the fermionic string rolls down towards the nonperturbative minimum of
the tachyon potential. This process represents the non-BPS brane decay
according to Sen's conjecture (see \cite{review-sft} for details). From
the point of view of the SFT the whole picture is not yet known and
only vacuum solutions were constructed. An effective field theory
description explaining the rolling tachyon in contrary is known and
numeric solutions describing the tachyon dynamics were
obtained~\cite{AJK}. This effective field theory description does
capture the nonlocality of the SFT. Linearizing the latter lagrangian
around the true vacuum one gets a model which is of main concern in the
present paper. The SFT inspired forms of function $\Fc(\Box_g)$, which
have the nonlocal operator $\exp(\alpha\Box_g)$, where $\alpha$ is a
constant, as a key ingredient, have been considered
in~\cite{AJV0701,AJV0711,MN}. Such functions have infinite number of
simple roots and maybe one double root.

The paper is organized as follows. In Section~2 we describe the
nonlocal SFT inspired cosmological model and its generalization. In
Section~3 we calculate the energy--momentum  tensor for different
special solutions. Using these formulae we build local actions and the
corresponding local Einstein equations. In Section 4 we propose the
algorithm to find particular solutions of the nonlocal Einstein
equations, solving only local ones, and prove the self-consistence of
it. Any solution for the obtained system of differential equations is a
particular solution for the initial nonlocal Einstein equations. In
Section~5 we summarize the obtained results and propose directions for
further investigations.


\section{Model setup}

The four-dimensional action with a quadratic potential, motivated
by the string field theory, has been studied
in~\cite{Koshelev07,AJV0701,AJV0711,MN,KV}. Such a model appears
as a linearization of the SFT inspired model in the neighborhood
of an extremum of the potential (see~\cite{KV} for details). For
linear models, solving the nonlocal equations using the technique,
proposed in~\cite{AJV0711}, is completely equivalent to solving
the equations using the diffusion-like partial differential
equations~\cite{MN}. In~\cite{MN} it has been shown that to fix
the initial data for the partial differential equations one can
use the initial data of the local fields. By linearising a
nonlinear model about a particular field value, one is able to
specify initial data for nonlinear models, which he then evolves
into the full nonlinear regime using the diffusion-like
equation~\cite{MN}.

In this paper we study nonlocal cosmological models with a quadratic
potential, in other words, a linear nonlocal model, which can be
described by the following action:
\begin{equation}
S=\int d^4x\sqrt{-g}\alpha^{\prime}\left(\frac{R}{16\pi
G_N}+\frac1{2g_o^2}\phi\Fc(\Box_g)\phi-\Lambda\right),
\label{action_model2}
\end{equation}
where $G_N$ is the Newtonian constant: $8\pi G_N=1/M_P^2$, where $M_P$
is the Planck mass, $\alpha^{\prime}$ is the string length squared (we
do not assume $\alpha^{\prime}=1/M_P^2$), $g_o$ is the string coupling
constant. We use the signature $(-,+,+,+)$, $g_{\mu\nu}$ is the metric
tensor, $R$ is the scalar curvature, $\Lambda$ is the cosmological
constant.

The function $\Fc$ is assumed to be an analytic function, therefore,
one can represent it by the convergent series expansion:
\begin{equation}
\Fc=\sum\limits_{n=0}^{\infty}f_n\Box_g^{\;n}.
\end{equation}
The function $\Fc$ may have infinitely many roots manifestly producing
thereby the nonlocality~\cite{noghosts,KV}.

This model has been studied in~\cite{Koshelev07,KV} with an additional
condition that all roots of the function $\Fc$ are simple. At the same
time the obtained formulae for the nonlocal energy--momentum tensor
(formulae (4.1) in~\cite{Koshelev07}) are valid in the case of multiple
roots as well and we use them in this paper.

In~\cite{AJV0701,AJV0711} the special class of functions $\Fc(\Box_g)$:
\begin{equation}
\label{F_SFT} \Fc_{sft}(\Box_g)={}-\xi^2 \Box_g+1-c\:e^{-2\Box_g},
\end{equation}
where $\xi$ is a real parameter and $c$ is a positive constant has been
considered. The action with $\Fc_{sft}(\Box_g)$ is interesting in
context of the SFT inspired models. In~\cite{MN} the model has been
generalized and a linear term has been added to the action.

 The function $\Fc_{sft}(\Box_g)$ has a double root if and only if
\begin{equation}
\label{c-xi} c=\frac{\xi^2}{2e}e^{2/\xi^2}.
\end{equation}
The double root $\tilde{J}_0$ is as follows
\begin{equation}
\label{z-0} \tilde{J}_0=\frac{1}{\xi^2}-\frac12.
\end{equation}

At any $\xi$ and $c$, which satisfy (\ref{c-xi}), the function
$\Fc_{sft}(J)$ has one and only one double root $\tilde{J}_0$ and
$\Fc_{sft}''(\tilde{J}_0)\neq 0$.

In this paper we consider in detail the case of an arbitrary analytic
function $\Fc$ with both double and simple roots.

To clarify the interest to consider the case of double roots let us
study a trivial example with
\begin{equation}
\Fc(\Box_g)=(\Box_g-J_1)(\Box_g-J_2).
\end{equation}

In the Minkowski space-time for $\phi$, depending only on time, we
obtain the following equation of motion
\begin{equation}
(\partial_t^2-J_1)(\partial_t^2-J_2)\phi(t)=0.
\end{equation}

This fourth-order differential equation can be written in the form of
system of two second order equations:
\begin{equation}
(\partial_t^2-J_1)\xi(t)=0,\qquad (\partial_t^2-J_2)\phi(t)=\xi(t).
\end{equation}

The first equation has the general solution
\begin{equation}
\xi(t)=C_1e^{\sqrt{J_1}t}+C_2e^{-\sqrt{J_1}t},
\end{equation}
where $C_1$ and $C_2$ are constants. So, we get the following second
order equation for $\phi$
\begin{equation}
(\partial_t^2-J_2)\phi(t)=C_1e^{\sqrt{J_1}t}+C_2e^{-\sqrt{J_1}t}.
\end{equation}
In the non-resonance case (two simple roots $J_1$ and $J_2$) we get the
following general solution
\begin{equation}
\phi(t)=\tilde{C}_1e^{\sqrt{J_1}t}+\tilde{C}_2e^{-\sqrt{J_1}t}+
\tilde{C}_3e^{\sqrt{J_2}t}+\tilde{C}_4e^{-\sqrt{J_2}t},
\end{equation}
whereas in the resonance case (one double root $J_2=J_1$) the general
solution is
\begin{equation}
\phi(t)=\tilde{C}_1e^{\sqrt{J_1}t}+
\tilde{C}_2e^{-\sqrt{J_1}t}+\tilde{C}_3te^{\sqrt{J_1}t}+\tilde{C}_4te^{-\sqrt{J_1}t},
\end{equation}
where $\tilde{C_k}$ are arbitrary constants. This trivial example shows that
behaviors of solutions in the cases of one double root and two simple
roots are essentially different and one can not approximate double
roots by two simple roots, which are at a very small distance.
Resonance phenomenons are important and actively studied in various
domains of physics.

\section{Energy--momentum tensor}

\subsection{The Einstein equations and energy--momentum tensor}

From action (\ref{action_model2}) we obtain the following equations
\begin{eqnarray}
&&G_{\mu\nu}=\frac{8\pi G_N}{g_o^2}T_{\mu\nu}-8\pi G_N\Lambda
g_{\mu\nu},
\label{EOJ_g}\\
&&\Fc(\Box_g)\phi=0, \label{EOJ_tau}
\end{eqnarray}
where $G_{\mu\nu}$ is the Einstein tensor, $T_{\mu\nu}$ is the
energy--momentum (stress) tensor~\cite{Koshelev07,KV}:
\begin{eqnarray}
\label{Tmunu}
T_{\mu\nu}&=&\frac{{}-2g_o^2}{\sqrt{-g}}\frac{\delta{S}}{\delta
g^{\mu\nu}}=\frac{1}{2}\sum_{n=1}^\infty
f_n\sum_{l=0}^{n-1}\left(\pd_\mu\Box_g^l\phi\pd_\nu\Box_g^{n-1-l}\phi
+\pd_\nu\Box_g^l\phi\pd_\mu\Box_g^{n-1-l}\phi-{}\right.\nonumber\\
&&\qquad\qquad\left.{}-g_{\mu\nu}\left(g^{\rho\sigma}
\pd_\rho\Box_g^l\phi\pd_\sigma\Box_g^{n-1-l}\phi+\Box_g^l\phi\Box_g^{n-l}\phi\right)\right),
\end{eqnarray}
\begin{equation}
\Box_g\equiv\frac1{\sqrt{-g}}\partial_{\mu}\sqrt{-g}g^{\mu\nu}\partial_{\nu}.
\end{equation}

 It is easy to check that the Bianchi identity is satisfied
on-shell and in a simple case $\Fc=f_1\Box_g+f_0$ the usual
energy--momentum tensor for the massive scalar field is reproduced.
Note that equation (\ref{EOJ_tau}) is an independent equation
consistent with system (\ref{EOJ_g}) due to the Bianchi identity.

In an arbitrary metric the energy--momentum tensor  (\ref{Tmunu}) can
be presented in the following form:
\begin{equation}
\label{TEV}
T_{\mu\nu}=E_{\mu\nu}+E_{\nu\mu}-g_{\mu\nu}\left(g^{\rho\sigma}
E_{\rho\sigma}+V\right),
\end{equation}
where
\begin{equation}
E_{\mu\nu}\equiv\frac{1}{2}\sum_{n=1}^\infty
f_n\sum_{l=0}^{n-1}\pd_\mu\Box_g^l\phi\pd_\nu\Box_g^{n-1-l}\phi,
\end{equation}
\begin{equation}
V\equiv\frac{1}{2}\sum_{n=1}^\infty
f_n\sum_{l=0}^{n-1}\Box_g^l\phi\Box_g^{n-l}\phi.
\end{equation}

\subsection{Energy--momentum tensor for special solutions}

Classical solutions to system (\ref{EOJ_g})--(\ref{EOJ_tau}) were
studied and analyzed in \cite{Koshelev07,AJV0701,AJV0711,KV}. The main
idea of finding the solutions to the equations of motion is to start
with equation (\ref{EOJ_tau}) and to solve it, assuming the function
$\phi$ is an eigenfunction of the d'Alembertian operator $\Box_g$. If
$\Box_g\phi=J\phi$, then such a function $\phi$ is a solution to
(\ref{EOJ_tau}) if and only if
\begin{equation}
\label{Betaequ} \Fc(J)=0.
\end{equation}
The latter condition is known as the \textit{characteristic} equation.
Note that values of roots of $\Fc(J)$ do not depend on the metric.

Let us denote simple roots of $\Fc$ as $J_i$ and double roots of
$\Fc$ as $\tilde{J}_k$.
 A particular solution of equation (\ref{EOJ_tau}) we  seek in the
 following form
 \begin{equation}
 \label{phi0}
 \phi_0=\sum\limits_{i=1}^{N_1}\phi_i+\sum\limits_{k=1}^{N_2}\tilde\phi_k,
\end{equation}
where
\begin{equation}
(\Box_g-J_i)\phi_i=0, \qquad (\Box_g-\tilde{J}_k)^2\tilde\phi_k=0.
\label{equphi}
\end{equation}

Without loss of generality we assume that for any $i_1$ and $i_2\neq
i_1$ conditions $J_{i_1}\neq J_{i_2}$ and
${\tilde{J}}_{i_1}\neq{\tilde{J}}_{i_2}$ are satisfied. Indeed, if, for
example, sum (\ref{phi0}) includes two summands $\phi_{i_{1}}$ and
$\phi_{i_{2}}$, which correspond to one and the same $J_i$, then we can
consider them as one summand $\phi_i\equiv \phi_{i_{1}}+\phi_{i_{2}}$,
which corresponds to $J_i$.

In previous papers~\cite{Koshelev07,AJV0701,AJV0711,MN,KV} only the
case of simple roots has been studied. In our paper we generalize this
analysis on  double roots. Our first goal is to calculate the
energy--momentum tensor for $\phi_0$. To obtain the general formula we
begin from a few particular cases.  Hereafter we denote the
energy--momentum tensor for the function $\phi(t)$ as
$T_{\mu\nu}(\phi)$.

\subsection{Simple roots}

If we have one simple root $\phi_1$ such that $\Box_g\phi_1=J_1\phi_1$,
then
\begin{equation}
E_{\mu\nu}(\phi_1)=\frac{1}{2}\sum_{n=1}^\infty
f_n\sum_{l=0}^{n-1}J_1^{n-1}\pd_{\mu}\phi_1\pd_{\nu}\phi_1
=\frac{{\Fc'(J_1)}}{2}\pd_{\mu}\phi_1\pd_{\nu}\phi_1.
\end{equation}
\begin{equation}
V(\phi_1)=\frac{1}{2}\sum_{n=1}^\infty
f_n\sum_{l=0}^{n-1}J_1^{n}\phi_1^2=\frac{J_1}{2}\sum_{n=1}^\infty
f_nnJ_1^{n-1}\phi_1^2=\frac{J_1\Fc'(J_1)}{2}\phi_1^2,
\end{equation}
where $\Fc'\equiv \frac{dF}{dJ}$.

In the case of two simple roots $\phi_1$ and $\phi_2$ we have
\begin{equation}
E_{\mu\nu}(\phi_1+\phi_2)=E_{\mu\nu}(\phi_1)+E_{\mu\nu}(\phi_2)+
E_{\mu\nu}^{cr}(\phi_1,\phi_2),
\end{equation}
where the cross term
\begin{equation}
E_{\mu\nu}^{cr}(\phi_1,\phi_2)=A_1\pd_{\mu}\phi_1\pd_{\nu}\phi_2+A_2\pd_{\mu}\phi_2\pd_{\nu}\phi_1.
\end{equation}

It is easy to calculate that
\begin{equation}
A_1=\frac{1}{2}\sum_{n=1}^\infty
f_nJ_1^{n-1}\sum_{l=0}^{n-1}\left(\frac{J_2}{J_1}\right)^l=\frac{\Fc(J_1)-\Fc(J_2)}{2(J_1-J_2)}=0,
\end{equation}
and
\begin{equation}
A_2=0.
\end{equation}
So, the cross term $E_{\mu\nu}^{cr}(\phi_1,\phi_2)=0$ and
\begin{equation}
E_{\mu\nu}(\phi_1+\phi_2)=E_{\mu\nu}(\phi_1)+E_{\mu\nu}(\phi_2).
\end{equation}
Similar calculations show
\begin{equation}
V(\phi_1+\phi_2)=V(\phi_1)+V(\phi_2).
\end{equation}

In the case of $N$ simple roots the following formula has been
obtained~\cite{KV} (see also~\cite{AJV0711}):
\begin{equation}
T_{\mu\nu}=\sum_{k=1}^N\Fc'(J_k)\left(\pd_\mu\phi_k\pd_\nu\phi_k
-\frac{1}{2}g_{\mu\nu}\left(g^{\rho\sigma}\pd_\rho\phi_k\pd_\sigma\phi_k+J_k\phi_k^2\right)
\right). \label{EOJ_g_onshell}
\end{equation}
Note that the last formula is exactly the energy--momentum tensor of
many free massive scalar fields. If $\Fc(J)$ has simple real roots,
then positive and negative values of $\Fc'(J_i)$ alternate, so we can
obtain phantom fields.

\subsection{One double root}

Let us consider the case, when all roots of $F(J)$, but one, are simple
and the last root is a double root. As we mentioned above this case is
interesting in context of the SFT inspired models.

Let $\tilde{J_1}$ is a double root. The fourth order differential
equation
\begin{equation}
(\Box_g-\tilde{J_1})(\Box_g-\tilde{J_1})\tilde\phi_1=0
\end{equation}
is equivalent to the following system of equations:
\begin{equation}
(\Box_g-\tilde{J_1})\tilde\phi_1=\varphi_1,\qquad
(\Box_g-\tilde{J_1})\varphi_1=0.
\end{equation}

It is convenient to write $\Box_g^l\tilde\phi_1$ in terms of
$\tilde\phi_1$ and $\varphi_1$:
\begin{equation}
\Box_g^l
\tilde\phi_1=\tilde{J}_1^l\tilde\phi_1+l\tilde{J}_1^{l-1}\varphi_1.
\label{Boxl_doubleroot}
\end{equation}

Using (\ref{Boxl_doubleroot}) we obtain
\begin{equation}
E_{\mu\nu}(\tilde\phi_1)=B_1\pd_\mu\tilde\phi_1\pd_\nu\tilde\phi_1+
B_2\pd_\mu\tilde\phi_1\pd_\nu\varphi_1+B_3\pd_\nu\tilde\phi_1\pd_\mu\varphi_1+
B_4\pd_\mu\varphi_1\pd_\nu\varphi_1,
\end{equation}
where
\begin{equation}
B_1=\frac{1}{2}\sum_{n=1}^\infty
f_n\sum_{l=0}^{n-1}\tilde{J}_1^{n-1}=\frac{1}{2}\sum_{n=1}^\infty
f_nn\tilde{J}_1^{n-1}=\frac{{\Fc'(\tilde{J}_1)}}{2}=0.
\end{equation}
\begin{equation}
B_2=\frac{1}{2}\sum_{n=1}^\infty
f_n\sum_{l=0}^{n-1}(n-l-1)\tilde{J}_1^{n-2}=\frac{1}{4}\sum_{n=1}^\infty
f_nn(n-1)\tilde{J}_1^{n-2}=\frac{{\Fc''(\tilde{J}_1)}}{4}.
\end{equation}
\begin{equation}
B_3=\frac{1}{2}\sum_{n=1}^\infty
f_n\sum_{l=0}^{n-1}l\tilde{J}_1^{n-2}=\frac{{\Fc''(\tilde{J}_1)}}{4}=B_2.
\end{equation}
\begin{equation}
B_4=\frac{1}{2}\sum_{n=1}^\infty
f_n\sum_{l=0}^{n-1}(n-l-1)l\tilde{J}_1^{n-3}=
\frac{1}{12}\sum_{n=1}^\infty
n(n-1)(n-2)f_n\tilde{J}_1^{n-3}=\frac{\Fc'''(\tilde{J}_1)}{12}.
\end{equation}

We have used the well-known formulae:
\begin{equation}
\sum_{l=0}^{n-1}l=\frac{n(n-1)}{2} \qquad\mbox{and}\qquad
\sum_{l=0}^{n-1}l^2=\frac{n(n-1)(2n-1)}{6}.
\end{equation}

Similar calculations give
\begin{equation}
V(\tilde{\phi}_1)=C_{1}\tilde\phi_1^2+C_{2}\tilde\phi_1\varphi_1+C_3\varphi_1^2,
\end{equation}
where
\begin{equation}
 C_1=\frac{1}{2}\sum_{n=1}^\infty
f_n\sum_{l=0}^{n-1}\tilde{J}_1^{n}=\frac{\tilde{J}_1}{2}\sum_{n=1}^\infty
f_nn\tilde{J}_1^{n-1}=\frac{\tilde{J}_1\Fc'(\tilde{J}_1)}{2}=0,
\end{equation}
\begin{equation}
C_2=\frac{1}{2}\sum_{n=1}^\infty
f_n\sum_{l=0}^{n-1}n\tilde{J}_1^{n-1}=\frac{{\tilde{J}_1\Fc''(\tilde{J}_1)}}{2}+\frac{{\Fc'(\tilde{J}_1)}}{2}=\frac{\tilde{J}_1\Fc''(\tilde{J}_1)}{2},
\end{equation}
\begin{equation}
C_3=\frac{1}{2}\sum_{n=1}^\infty
f_n\sum_{l=0}^{n-1}l(n-l)\tilde{J}_1^{n-2}=\frac{{\tilde{J}_1\Fc'''(\tilde{J}_1)}}{12}+\frac{{\Fc''(\tilde{J}_1)}}{4}.
\end{equation}

Thus, for one double root we obtain the following result:
\begin{equation}
\label{Edr} E_{\mu\nu}(\tilde\phi_1)=
\frac{{\Fc''(\tilde{J}_1)}}{4}(\pd_\mu\tilde\phi_1\pd_\nu\varphi_1+\pd_\nu\tilde\phi_1\pd_\mu\varphi_1)+
\frac{\Fc'''(\tilde{J}_1)}{12}\pd_\mu\varphi_1\pd_\nu\varphi_1,
\end{equation}
\begin{equation}
\label{Vdr}
V(\tilde{\phi_1})=\frac{\tilde{J}_1\Fc''(\tilde{J}_1)}{2}\tilde\phi_1\varphi_1+
\left(\frac{{\tilde{J}_1\Fc'''(\tilde{J}_1)}}{12}+\frac{{\Fc''(\tilde{J}_1)}}{4}\right)\varphi_1^2.
\end{equation}

For one simple root $J_2$ (the function $\phi_2$ satisfies the equation
$\Box_g\phi_2=J_2\phi_2$) and one double root $\tilde J_1$ we obtain:
\begin{equation}
E_{\mu\nu}(\tilde\phi_1+\phi_2)=E_{\mu\nu}(\tilde\phi_1)+E_{\mu\nu}(\phi_2)+
E_{\mu\nu}^{cr}(\tilde\phi_1,\phi_2),
\end{equation}
where
\begin{equation}
E_{\mu\nu}^{cr}(\tilde\phi_1,\phi_2)=B_5\pd_\mu\tilde\phi_1\pd_\nu\phi_2+
B_6\pd_\nu\tilde\phi_1\pd_\mu\phi_2+B_7\pd_\mu\varphi_1\pd_\nu\phi_2+
B_8\pd_\nu\varphi_1\pd_\mu\phi_2.
\end{equation}

It is easy to calculate:
\begin{equation}
B_5=\frac{1}{2}\sum_{n=1}^\infty
f_nJ_2^{n-1}\sum_{l=0}^{n-1}\left(\frac{\tilde{J}_1}{J_2}\right)^l=\frac{\Fc(J_2)-\Fc(\tilde{J}_1)}{2(J_2-\tilde{J}_1)}=0.
\end{equation}
\begin{equation}
B_6=0.
\end{equation}

To calculate
\begin{equation}
B_7=\frac{1}{2}\sum_{n=1}^\infty
f_nJ_2^{n}\sum_{l=0}^{n-1}l\left(\frac{\tilde{J}_1}{J_2}\right)^l
\end{equation}
 we use
\begin{equation}
\sum_{l=0}^{n-1}ly^{l-1}=\frac{d}{dy}\sum_{l=0}^{n-1}y^{l}=\frac{d}{dy}\left(\frac{1-y^n}{1-y}\right)=
\frac{(n-1)y^n-ny^{n-1}+1}{(1-y)^2}
\end{equation}
and obtain
\begin{equation}
B_7=\frac{J_2^2}{2(J_2-\tilde{J}_1)}\Fc'(\tilde{J}_1)+\frac{J_2^2}{2(J_2-\tilde{J}_1)^2}(\Fc(\tilde{J}_1)+\Fc(J_2))=0.
\end{equation}

Similar calculations give
\begin{equation}
B_8=0.
\end{equation}

It is easy to obtain that
\begin{equation}
V(\tilde\phi_1+\phi_2)=\frac{1}{2}\sum_{n=1}^\infty
f_n\sum_{l=0}^{n-1}\Box_g^l(\tilde\phi_1+\phi_2)\Box_g^{n-l}(\tilde\phi_1+\phi_2)=
V(\tilde\phi_1)+V(\phi_2).
\end{equation}

The calculations are straightforwardly generalized on the case of one
double root and an arbitrary number of simple roots. Therefore, we
obtain the following formula
\begin{equation}
T_{\mu\nu}\left(\tilde\phi_1+\sum_{k=1}^N\phi_k\right)=
T_{\mu\nu}\left(\tilde\phi_1\right)+T_{\mu\nu}\left(\sum_{k=1}^N\phi_k\right),
\end{equation}
where
\begin{equation}
T_{\mu\nu}(\tilde\phi_1)=E_{\mu\nu}(\tilde\phi_1)+E_{\nu\mu}(\tilde\phi_1)
-g_{\mu\nu}\left(g^{\rho\sigma}
E_{\rho\sigma}(\tilde\phi_1)+V(\tilde\phi_1)\right).
\end{equation}

So,  we conclude that in the case of one double root the
energy--momentum tensor can be separated into energy--momentum tensors
for different modes of nonlocal scalar field, which correspond to
different roots of $\Fc$.

\subsection{The general formulae}

Let us consider the case of two double roots $\tilde{J}_1$ and
$\tilde{J}_2$. We can write
\begin{equation}
E_{\mu\nu}(\tilde\phi_1+\tilde\phi_2)=E_{\mu\nu}(\tilde\phi_1)+E_{\mu\nu}(\tilde\phi_2)+
E_{\mu\nu}^{cr}(\tilde\phi_1,\tilde\phi_2),
\end{equation}
where
\begin{equation}
\begin{array}{rcl}
E_{\mu\nu}^{cr}(\tilde\phi_1,\tilde\phi_2)&=&B_{10}\pd_\mu\tilde\phi_1\pd_\nu\tilde\phi_2+
B_{11}\pd_\nu\tilde\phi_1\pd_\mu\tilde\phi_2+B_{12}\pd_\mu\tilde\phi_1\pd_\nu\varphi_2+
{}\\&+&B_{13}\pd_\nu\tilde\phi_1\pd_\mu\varphi_2+B_{14}\pd_\mu\varphi_1\pd_\nu\tilde\phi_2+
B_{15}\pd_\nu\varphi_1\pd_\mu\tilde\phi_2+{}\\&+&B_{16}\pd_\mu\varphi_1\pd_\nu\varphi_2+
B_{17}\pd_\nu\varphi_1\pd_\mu\varphi_2.
\end{array}
\end{equation}
Using computations, which are similar to computations of $B_5$ and
$B_7$, it is easy to see
\begin{equation}
B_{10}=B_{11}=B_{12}=B_{13}=B_{14}=B_{15}=0.
\end{equation}

It is suitable to present
\begin{equation}
B_{16}=\frac{1}{2}\sum_{n=1}^\infty
f_n\sum_{l=0}^{n-1}l(n-l-1)\tilde{J}_1^{l-1}\tilde{J}_2^{n-l-2}
\end{equation}
 in the following form:
\begin{equation}
B_{16}= \frac{1}{2}\sum_{n=1}^\infty f_n\tilde{J}_2^{n-1}
\sum_{l=0}^{n-1}(n-l-1)l\varpi^{l-1},
\end{equation}
where $\varpi\equiv\tilde{J}_1/\tilde{J}_2$. Using
\begin{equation*}
\sum_{l=0}^{n-1}(n-l-1)l\varpi^{l-1}=n\sum_{l=0}^{n-1}l\varpi^{l-1}-\varpi\sum_{l=0}^{n-1}(l-1)l\varpi^{l-2}=
n\frac{d}{d\varpi}\left(\sum_{l=0}^{n-1}\varpi^{l}\right)-\varpi
\frac{d^2}{d\varpi^2}\left(\sum_{l=0}^{n-1}\varpi^{l}\right)
\end{equation*}
and
\begin{equation}
\sum_{l=0}^{n-1}\varpi^{l}=\frac{1-\varpi^{n}}{1-\varpi},
\end{equation}
we obtain
\begin{equation}
 \sum_{l=0}^{n-1}(n-l-1)l\varpi^{l-1}=
n\frac{1+\varpi^{n-1}(2\varpi-1)}{(\varpi-1)^2}+
2\varpi\frac{1-\varpi^n}{(\varpi-1)^3}.
\label{symplify}
\end{equation}

Thus we get
\begin{equation}
B_{16}=\frac{\tilde{J}_2(\Fc(\tilde{J}_2)-\Fc(\tilde{J}_1))}{(\tilde{J}_2-\tilde{J}_1)^3}
+
\frac{\tilde{J}_2(2\tilde{J}_1-\tilde{J}_2)\Fc'(\tilde{J}_1)+\tilde{J}_2^2\Fc'(\tilde{J}_2)}{2(\tilde{J}_1-\tilde{J}_2)^2}.
\end{equation}

So, $B_{16}=0$. The similar calculations prove that $B_{17}=0$ and we
come to the following result:
\begin{equation}
E_{\mu\nu}(\tilde\phi_1+\tilde\phi_2)=E_{\mu\nu}(\tilde\phi_1)+E_{\mu\nu}(\tilde\phi_2).
\end{equation}
We also obtain
\begin{equation}
V(\tilde\phi_1+\tilde\phi_2)=V(\tilde\phi_1)+V(\tilde\phi_2).
\end{equation}

The results, obtained for two summands, can be straightforwardly
generalized on an arbitrary number of summands. So, we obtain that for
any analytical function $\Fc$, which has simple roots $J_i$ and double
roots $\tilde{J}_k$, and any $\phi_0$ given by (\ref{phi0}) the
energy--momentum tensor
\begin{equation}
T_{\mu\nu}\left(\phi_0\right)=
T_{\mu\nu}\left(\sum\limits_{i=1}^{N_1}\phi_i+\sum\limits_{k=1}^{N_2}\tilde\phi_k\right)=
\sum\limits_{i=1}^{N_1}T_{\mu\nu}(\phi_i)+\sum\limits_{k=1}^{N_2}T_{\mu\nu}(\tilde\phi_k).
\label{Tmunugen}
\end{equation}

The result has been obtained for an arbitrary metric $g_{\mu\nu}$.

Considering the following local action
\begin{equation}
\begin{array}{rl}
\displaystyle S_{loc}=&\displaystyle \int
d^4x\sqrt{-g}\left(\frac{R}{16\pi
G_N}-\Lambda-{}\right.\\
-&\displaystyle
\frac{1}{g_o^2}\left(\sum_{i=1}^{N_1}\frac{\Fc'(J_i)}{2}\left(g^{\mu\nu}\pd_\mu\phi_i\pd_\nu\phi_i
+J_i\phi_i^2\right)-{}\right.\\
-&\displaystyle
\sum_{k=1}^{N_2}\left(g^{\mu\nu}\left(\frac{{\Fc''(\tilde{J}_k)}}{2}\pd_\mu
\tilde{\phi}_k\pd_\nu\varphi_k+
\frac{\Fc'''(\tilde{J}_k)}{12}\pd_\mu\varphi_k\pd_\nu\varphi_k\right)+
\right.\\
+&\displaystyle \left.\left.\left.
\frac{\tilde{J}_k\Fc''(\tilde{J}_k)}{2}\tilde\phi_k\varphi_k
+\left(\frac{{\tilde{J}_k\Fc'''(\tilde{J}_k)}}{12}+\frac{{\Fc''(\tilde{J}_k)}}{4}\right)\varphi_k^2\right)\right)\right),
\label{Sloc}
\end{array}
\end{equation}
we can see that solutions of the system of the Einstein equations and
equations in $\phi_k$, $\tilde{\phi}_k$ and $\varphi_k$, obtained from
this action, solves the initial system of nonlocal equations
(\ref{EOJ_g}) and (\ref{EOJ_tau}). Thus, we obtained that special
solutions of nonlocal equations one can find solving system of local
(differential) equations.

To clarify physical interpretation we diagonalize the kinetic
terms of scalar fields $\tilde{\phi}_k$ and $\varphi_k$ in action
(\ref{Sloc}).  It is convenient to present (\ref{Sloc}) as
follows:
\begin{equation}
S_{loc}=\int\! d^4x\sqrt{-g}\left(\frac{R}{16\pi
G_N}-\Lambda\right)+\sum_{i=1}^{N_1}S_i+\sum_{k=1}^{N_2}\tilde{S}_k,
\end{equation}
where
\begin{equation}
S_i=\!{}-\frac{1}{2g_o^2}\int\!
d^4x\sqrt{-g}\Fc'(J_i)\left(g^{\mu\nu}\pd_\mu\phi_i\pd_\nu\phi_i
+J_i\phi_i^2\right),
\end{equation}

\begin{equation}
\begin{array}{rl}
\!\displaystyle\tilde{S}_k=&\!\displaystyle\! {}-\frac{1}{g_o^2}\int\!
d^4x\sqrt{-g}\left[g^{\mu\nu}\left(\frac{{\Fc''(\tilde{J}_k)}}{4}\left(\pd_\mu
\tilde{\phi}_k\pd_\nu\varphi_k+\pd_\nu
\tilde{\phi}_k\pd_\mu\varphi_k\right)+{}\right.\right.\\
\displaystyle +&\displaystyle
\left.\frac{\Fc'''(\tilde{J}_k)}{12}\pd_\mu\varphi_k\pd_\nu\varphi_k\right)+\\
\displaystyle +&\displaystyle \left.
\frac{\tilde{J}_k\Fc''(\tilde{J}_k)}{2}\tilde\phi_k\varphi_k
+\left(\frac{{\tilde{J}_k\Fc'''(\tilde{J}_k)}}{12}+\frac{{\Fc''(\tilde{J}_k)}}{4}\right)\varphi_k^2\right],
\label{Slocdr}
\end{array}
\end{equation}

Expressing $\tilde{\phi}_k$ and $\varphi_k$ in terms of new fields
$\xi_k$ and $\chi_k$:
\begin{equation}
    \tilde{\phi}_k=\frac{1}{2\Fc''(\tilde{J}_k)}\left((\Fc''(\tilde{J}_k)-\frac{1}{3}\Fc'''(\tilde{J}_k))
    \xi_k-(\Fc''(\tilde{J}_k)+\frac{1}{3}\Fc'''(\tilde{J}_k))\chi_k\right),
\end{equation}
\begin{equation}
    \varphi_k=\xi_k+\chi_k,
\end{equation}
we obtain the corresponding $\tilde{S}_k$ in the following form:
\begin{equation*}
\begin{array}{l}
\displaystyle\tilde{S}_k={}-\frac{1}{g_o^2}\int
d^4x\sqrt{-g}\left[g^{\mu\nu}\frac{\Fc''(\tilde{J}_k)}{4}(\pd_\mu
\xi_k\pd_\nu\xi_k-\pd_\nu
\chi_k\pd_\mu\chi_k)+{}\right.\\
\displaystyle
{}+\frac{\tilde{J}_k}{4}\left(\left(\Fc''(\tilde{J}_k)-\frac{1}{3}\Fc'''(\tilde{J}_k)\right)
    \xi_k-\left(\Fc''(\tilde{J}_k)+\frac{1}{3}\Fc'''(\tilde{J}_k)\right)\chi_k\right)(\xi_k+\chi_k)+{}\\
\displaystyle {}+ \left.
\left(\frac{{\tilde{J}_k\Fc'''(\tilde{J}_k)}}{12}+\frac{{\Fc''(\tilde{J}_k)}}{4}\right)(\xi_k+\chi_k)^2\right].
\end{array}
\end{equation*}

It is easy to see that each $\tilde{S}_k$ includes one phantom
scalar field and one standard scalar field. So, in the case of one
double root we obtain a quintom model. In the Minkowski space
appearance of  phantom fields in models, when $\Fc(\Box)$ has a
double root, has been obtained in~\cite{PaisU}.

\section{The algorithm of localization}

The obtained formulae allow us to seek particular solutions for
nonlocal gravitational models with quadratic potentials, which
described by action (\ref{action_model2}), in the following way:
\begin{itemize}

\item Find roots of the function $\Fc(J)$ and calculate orders of them.

\item Select an finite number of simple and double roots.

\item Construct the corresponding local action by formula (\ref{Sloc}).

\item Obtain a system of the Einstein equations and equations of
motion. The obtained system is a finite order system of differential
equations, in other words we get a local system.

\item Seek solutions of the obtained local system.

\end{itemize}

\textbf{Remark 1.} If $\Fc(J)$ has an infinity number of roots then one
nonlocal model corresponds to infinity number of different local
models. In this case the initial nonlocal action (\ref{action_model2})
generates infinity number of local actions (\ref{Sloc}).

\textbf{Remark 2.} We should prove that our algorithm is
self-consistent. To construct local action (\ref{Sloc}) we assume that
equations (\ref{equphi}) are satisfied. Therefore, our algorithm is
correct only if these equations can be obtained from the local action
(\ref{Sloc}). The straightforward calculations show that
\begin{equation}
\frac{\delta{S_{loc}}}{\delta \phi_i}=0 \qquad \Leftrightarrow \qquad
\Box_g\phi_i=J_i\phi_i,
\end{equation}
\begin{equation}
\frac{\delta{S_{loc}}}{\delta \tilde{\phi}_k}=0 \qquad \Leftrightarrow
\qquad \Box_g\varphi_k=\tilde{J}_k\varphi_k. \label{equvarphi}
\end{equation}
Using (\ref{equvarphi}) we obtain
\begin{equation}
\frac{\delta{S_{loc}}}{\delta \varphi_k}=0 \qquad \Leftrightarrow
\qquad \Box_g\tilde{\phi}_k=\tilde{J}_k\tilde{\phi}_k+\varphi_k.
\end{equation}

So, our way of localization is self-consistent in the case of $\Fc(J)$
with simple and double roots. The self-consistence of similar approach
for $\Fc(J)$ with only simple roots has been proven
in~\cite{AJV0711,KV}.

In spite of the above-mention equations we obtain  from $S_{loc}$ the
Einstein equations:
\begin{equation}
G_{\mu\nu}=\frac{8\pi G_N}{g_o^2}T_{\mu\nu}(\phi_0)-8\pi G_N\Lambda
g_{\mu\nu},
\end{equation}
where $\phi_0$ is given by (\ref{phi0}) and $T_{\mu\nu}(\phi_0)$ can be
calculated by (\ref{Tmunugen}).

So, we obtained such systems of differential equations that any
solutions of these systems are particular solutions of the initial
nonlocal equations (\ref{EOJ_g}) and (\ref{EOJ_tau}).

\section{Conclusion}

The main result of this paper is the explicit proof that nonlocal
cosmological model can be localized not only in the case, when
$\Fc(\Box_g)$ has only simple roots. We have found the way to find
particular solutions of the nonlocal Einstein equations in the
case when an analytic function $\Fc(\Box_g)$ has both simple and
double roots. We prove that the same functions solve the initial
nonlocal Einstein equations and the obtained local Einstein
equations. We have found the corresponding local actions and
proved the self-consistence of our approach. The result has been
obtained for an arbitrary metric, so it can be used not only to
find solutions in the Friedmann--Robertson--Walker metric, but
also to find other interesting solutions, for example, black hole
solutions. In the case of simple roots some exact solutions in the
Friedmann--Robertson--Walker metric have been found
in~\cite{AJV0711} (the stability of these solutions is considered
in~\cite{ABJV0903}).

Looking a step further it is interesting to consider nonlocal models
with an arbitrary analytic $\Fc(\Box_g)$, without any restrictions on
order of roots.  The consideration of simple roots in
papers~\cite{AJV0711,KV} and double roots in this paper allows us to
make the conjecture that the existence of local actions, which
correspond to a nonlocal action, does not depend on order of
$\Fc(\Box_g)$ roots and the method of finding particular solutions of
the nonlocal Einstein equations can be generalized on a nonlocal action
with an arbitrary analytic $\Fc(\Box_g)$.


\section*{Acknowledgements}

The author is grateful to I.Ya.~Aref'eva, A.S.~Koshelev and
A.F.~Zakharov for useful and stimulating discussions. This work is
supported in part by RFBR grant 08-01-00798, grant of Russian Ministry
of Education and Science of Russia NSh-1456.2008.2 and state contract
of Russian Federal Agency for Science and Innovations 02.740.11.5057.


\begin{thebibliography}{72}

\bibitem{review-sft}  K.~Ohmori,\textit{ A Review on Tachyon Condensation in Open String Field Theories},
\hepth{0102085};\\
 I.Ya.~Aref'eva, D.M.~Belov, A.A.~Giryavets, A.S.~Koshelev, and
P.B.~Medvedev, \textit{Noncommutative Field Theories and (Super)String Field Theories}, \hepth{0111208};\\
 W. Taylor, \textit{Lectures on D-branes, tachyon condensation, and string field theory},
 \hepth{0301094}
\bibitem{padic}
L.~Brekke, P.G.O.~Freund, M.~Olson, and E.~Witten,
\textit{Nonarchimedean
String Dynamics}, Nucl. Phys. \textbf{B 302} (1988) 365--402;\\
P.H. Frampton, Ya. Okada, \textit{Effective Scalar Field Theory of
$P$-Adic String},
Phys. Rev. \textbf{D 37} (1988) 3077--3079;\\
V.S.~Vladimirov, I.V.~Volovich, and E.I.~Zelenov, \textit{$p$-adic
Analysis and Mathe\-matical Physics}, WSP, Singapore,
1994;\\
B. Dragovich, A.Yu. Khrennikov, S.V. Kozyrev, and I.V. Volovich,
\textit{$p$-Adic Mathematical Physics}, Anal. Appl. \textbf{1} (2009)
1--17, \arXivid{0904.4205}
\bibitem{IA1} I.Ya. Aref'eva, \textit{Nonlocal String Tachyon as
a Model for Cosmological Dark Energy},  AIP Conf. Proc.
{\bf 826} (2006) 301--311, \astroph{0410443};\\
I.Ya.~Aref'eva, {\it D-brane as a Model for Cosmological Dark Energy},
in: "Contents and Structures of the Universe",
            eds. C. Magneville, R. Ansari, J. Dumarchez,
             and J.T.T. Van, \textit{Proc. of the XLIst Rencontres de
Moriond}, 2006, pp.~131--135;\\
I.Ya.~Aref'eva, \textit{Stringy Model of Cosmological Dark Energy}, AIP
Conf. Proc. \textbf{957} (2007) 297--300, \arXivid{0710.3017}
\bibitem{AJ} I.Ya. Aref'eva and  L.V. Joukovskaya,
\textit{Time Lumps in Nonlocal Stringy Models and Cosmological
Applications}, JHEP \textbf{0510} (2005) 087, \hepth{0504200}
\bibitem{Calcagni}  G. Calcagni, \textit{Cosmological tachyon from cubic string field theory},
 JHEP \textbf{0605} (2006) 012,  \hepth{0512259}
\bibitem{Barnaby} N. Barnaby, T. Biswas, and  J.M. Cline,
\textit{p-adic Inflation},  JHEP \textbf{0704} (2007) 056,
\hepth{0612230}\\
 N. Barnaby and  J.M. Cline, \textit{Large Nongaussianity from Nonlocal
Inflation}, JCAP \textbf{0707} (2007) 017,
\arXivid{0704.3426}\\
 N. Barnaby, \textit{Nonlocal Inflation}, Can. J. Phys. \textbf{87} (2009) 189--194,
\arXivid{0811.0814}
\bibitem{Koshelev07} A.S. Koshelev, \textit{Non-local SFT Tachyon and Cosmology},
JHEP \textbf{0704} (2007) 029, \hepth{0701103}
\bibitem{AJV0701} I.Ya. Aref'eva, L.V. Joukovskaya, and S.Yu. Vernov,
 \textit{Bouncing and accelerating solutions in nonlocal stringy models}
 JHEP \textbf{0707} (2007) 087, \hepth{0701184}
\bibitem{AVzeta}
 I.Ya. Aref'eva and  I.V.  Volovich,  \textit{Quantization of the Riemann
 Zeta-Function and Cosmology},  Int. J. of Geom. Meth.
Mod. Phys.  \textbf{4} (2007) 881--895, \hepth{0701284}
\bibitem{Lidsey07} J.E. Lidsey,  \textit{Stretching the Inflaton
Potential with Kinetic Energy},  Phys. Rev.  {\bf D 76} (2007) 043511,
\hepth{0703007}
\bibitem{Calcagni07}
 G. Calcagni, M. Montobbio, and  G. Nardelli, \textit{A route to
nonlocal cosmology}, Phys. Rev. \textbf{D 76} (2007) 126001,
\arXivid{0705.3043}\\
G. Calcagni  and G. Nardelli,\textit{ Tachyon solutions in boundary and
cubic string field theory}, Phys. Rev. \textbf{D 78} (2008) 126010, \arXivid{0708.0366};\\
G. Calcagni, M. Montobbio, and  G. Nardelli, \textit{Localization of
nonlocal theories},
Phys. Lett. \textbf{B 662} (2008) 285--289, \arXivid{0712.2237};\\
G. Calcagni and  G. Nardelli, \textit{Nonlocal instantons and solitons
in string models},  Phys. Lett. \textbf{B 669} (2008) 102--112,
\arXivid{0802.4395}

 \bibitem{LJ-PR} L.V. Joukovskaya, {\it Dynamics in nonlocal cosmological models derived from string field theory}
  Phys. Rev. \textbf{D 76} (2007) 105007, \arXivid{0707.1545};\\
   L.V. Joukovskaya, {\it Rolling tachyon in nonlocal cosmology} AIP
Conf. Proc. \textbf{957} (2007) 325--328,  \arXivid{0710.0404};\\
 L.V. Joukovskaya, {\it  Dynamics with Infinitely Many Time Derivatives in Friedmann--Robertson--Walker Background
 and Rolling Tachyon},
  JHEP \textbf{0902} (2009) 045, \arXivid{0807.2065}

\bibitem{noghosts}
 N. Barnaby and  N. Kamran,  \textit{Dynamics
with Infinitely Many Derivatives: The Initial Value Problem}, JHEP {\bf
0802} (2008) 008, \arXivid{0709.3968}

\bibitem{AJV0711} I.Ya. Aref'eva, L.V. Joukovskaya, and S.Yu. Vernov,
 \textit{Dynamics in nonlocal linear models in the Friedmann--Robertson--Walker metric},
 J. Phys. A: Math. Theor. \textbf{41} (2008) 304003,
 \arXivid{0711.1364}

\bibitem{MN}  D.J. Mulryne and N.J. Nunes,   \textit{Diffusing non-local inflation: Solving
the field equations as an initial value problem},
Phys. Rev. \textbf{D 78} (2008) 063519, \arXivid{0805.0449};\\
D.J. Mulryne and N.J. Nunes, \textit{Non-linear non-local Cosmology},
AIP Conf. Proc. \textbf{1115} (2009) 329--334, \arXivid{0810.5471}

\bibitem{BarnabyKamran} N. Barnaby and
N. Kamran,  \textit{Dynamics with Infinitely Many Derivatives: Variable
Coefficient Equations}, JHEP {\bf 0812} (2008) 022, \arXivid{0809.4513}


\bibitem{KV}
A.S. Koshelev and S.Yu. Vernov, \textit{Cosmological perturbations in
SFT inspired non-local scalar field models}, \arXivid{0903.5176}

\bibitem{CN}
G. Calcagni and G. Nardelli, \textit{Cosmological rolling solutions of
nonlocal theories},  Int. J. Mod. Phys. \textbf{D 19} (2010) 329,
\arXivid{0904.4245}


\bibitem{Hawking-Ellis} S.W.~Hawking and G.F.R.~Ellis, \textit{The
  Large Scale Structure of space-time}, Cambridge University Press,
  Cambridge, 1973.

\bibitem{cyclic}
P.J.~Steinhardt and N.~Turok, \textit{Cosmic evolution in a cyclic
universe}, Phys. Rev. {\bf D 65} (2002) 126003, \hepth{0111098}

\bibitem{GV} M. Gasperini and G. Veneziano, {\it The Pre-big bang scenario in string
cosmology}, Phys. Rept. \textbf{373} (2003) 1--212, \hepth{0207130}

\bibitem{Caldwell} R.R.~Caldwell,  \textit{A Phantom Menace? Cosmological consequences of a
dark energy component with super-negative equation of state}, Phys.
Lett.  \textbf{B 545} (2002) 23--29, \astroph{9908168}

\bibitem{0301273} S.M. Carroll, M. Hoffman, and M. Trodden,
{\it Can the dark energy equation-of-state parameter w be less than
$-1$?}, Phys. Rev.  \textbf{D 68} (2003) 023509, \astroph{0301273}

\bibitem{0311312}
  J.M.~Cline, S.~Jeon, and G.D.~Moore,
  {\it The phantom menaced: Constraints on low-energy effective ghosts},
  Phys. Rev. {\bf D 70} (2004) 043543,
 \hepth{0311312}.

\bibitem{0406043}
  S.D.H.~Hsu, A.~Jenkins, and M.B.~Wise,
  {\it Gradient instability for} $w < -1$,
  Phys. Lett.  {\bf B 597} (2004)  270--274, \astroph{0406043};\\
R.V. Buniy, S.D.H. Hsu, and  B.M. Murray, {\it The null energy
condition and instability}, Phys. Rev. {\bf D 74} (2006) 063518,
\hepth{0606091}

\bibitem{McInnes} B. McInnes,  {\it The Phantom divide in string gas cosmology},
Nucl.Phys. \textbf{B 718} (2005) 55--82, \hepth{0502209}




\bibitem{GWG}
 G.W. Gibbons, \textit{Phantom Matter and the Cosmological Constant},
 \hepth{0302199}

\bibitem{NojiriOdintsov}
 S. Nojiri and S.D. Odintsov,  \textit{Quantum deSitter cosmology and phantom matter},
   Phys. Lett.  {\bf B 562} (2003) 147--152, \hepth{0303117};\\
 S. Nojiri and S.D. Odintsov, \textit{deSitter brane universeinduced by phantom and quantum effects},
 Phys. Lett. {\bf B 565} (2003) 1--9, \hepth{0304131}

\bibitem{Caldwell03}
 R.R. Caldwell, M. Kamionkowski,  and  N.N. Weinberg, \textit{Phantom Energy and Cosmic Doomsday},
 Phys. Rev. Lett. \textbf{91} (2003) 071301; \astroph{0302506}

\bibitem{Woodard} V.K. Onemli  and   R.P. Woodard, \textit{Super-Acceleration
from Massless, Minimally Coupled $\varphi^4$}, Class. Quant. Grav.
\textbf{19} (2002) 4607--4626; \grqc{0204065}; \\
 V.K. Onemli  and   R.P. Woodard,  \textit{Quantum effects can render $w<-1$
 on cosmological scales}, Phys. Rev. \textbf{D 70} (2004) 107301,
 \grqc{0406098}

\bibitem{Carroll} S.M. Carroll, A. De Felice, and  M. Trodden, \textit{Can we be tricked into thinking that
$w$ is less than $-1$?}, Phys.  Rev. \textbf{D 71} (2005) 023525,
\astroph{0408081}


\bibitem{AV-NEC}
I.Ya.~Aref'eva and I.V.~Volovich, \textit{On the null energy condition
and cosmology}, Theor. Math. Phys. \textbf{155} (2008)  503--511 [Teor.
Mat. Fiz. \textbf{155} (2008) 3--12], \hepth{0612098}

\bibitem{RAS}
R.~Kallosh, J.U. Kang, A.~Linde, and V.~Mukhanov, {\it The New
Ekpyrotic Ghost}, JCAP {\bf 0804} (2008) 018, \arXivid{0712.2040}

\bibitem{SW}
  S.~Weinberg,
  \textit{Effective Field Theory for Inflation},
  Phys. Rev.  {\bf D  77} (2008) 123541,
  \arXivid{0804.4291};\\
  J.Z.~Simon,
  \textit{Higher derivative Lagrangians, non-locality, problems and
  solutions},
  Phys. Rev.  \textbf{D 41} (1990) 3720--3733

\bibitem{Creminelli0812} P. Creminelli, G. D'Amico, J. Norena, and  F.
Vernizzi, \textit{The Effective Theory of Quintessence: the $w<-1$ Side
Unveiled}, JCAP \textbf{0902} (2009) 018, \arXivid{0811.0827}

\bibitem{data}
 A.G. Riess \textit{et al.}  [Supernova Search Team collaboration],
\textit{Type Ia Supernova Discoveries at $z>1$ From the Hubble Space
Telescope: Evidence for Past Deceleration and Constraints on Dark
Energy Evolution},  Astrophys. J. \textbf{607} (2004) 665--687, \astroph{0402512};\\
 M. Tegmark \textit{et al.} [SDSS collaboration], \textit{The 3D power
spectrum of galaxies from the SDS},  Astroph.~J. \textbf{606} (2004) 702--740, \astroph{0310725};\\
 P. Astier {\it et al.},  \textit{The Supernova Legacy Survey: Measurement
of $\Omega_M$, $\Omega_\Lambda$ and $w$ from the First Year Data Set},
Astron. Astrophys. \textbf{447} (2006) 31--48,
\astroph{0510447};\\
Shirley Ho, Chr. M. Hirata, N.  Padmanabhan, U. Seljak, and N. Bahcall,
\textit{Correlation of CMB with large-scale structure: I. ISW
Tomography and Cosmological Implications},
Phys. Rev. \textbf{D 78} (2008) 043519,   \arXivid{0801.0642};\\
 W.M. Wood-Vasey {\it et al.} [ESSENCE Collaboration], \textit{Observational
Constraints on the Nature of the Dark Energy: First Cosmological
Results from the ESSENCE Supernova Survey},  Astrophys. J. \textbf{666}
(2007) 694--715, \astroph{0701041};\\
D.~Baumann {\it et al.}  [CMBPol Study Team Collaboration],
  \textit{CMBPol Mission Concept Study: Probing Inflation with CMB Polarization},
  AIP Conf. Proc. \textbf{1141} (2009) 10--120, \arXivid{0811.3919}\\
E. Komatsu, J. Dunkley, M.R. Nolta, C.L. Bennett, B. Gold, G. Hinshaw,
N. Jarosik, D.~Larson, M. Limon, L. Page, D.N. Spergel, M. Halpern,
R.S. Hill, A. Kogut, S.S.~Meyer, G.S. Tucker, J.L. Weiland, E. Wollack,
and E.L. Wright, \textit{Five-Year Wilkinson Microwave Anisotropy Probe
(WMAP) Observations: Cosmological Interpretation}, {Astrophys.\ J.\
Suppl.\ } \textbf{180} (2009) {330--376},
\arXivid{0803.0547}; \\
M. Kilbinger, K. Benabed, J. Guy, P. Astier, I. Tereno, L. Fu, D.
Wraith, J. Coupon, Y.~Mellier, C. Balland, F.R. Bouchet, T. Hamana, D.
Hardin, H.J. McCracken, R.~Pain, N. Regnault, M. Schultheis, and H.
Yahagi, \textit{Dark energy constraints and correlations with
systematics from CFHTLS weak lensing, SNLS supernovae Ia and WMAP5},
Astron. Astrophys.  \textbf{497} (2009) 677--688, \arXivid{0810.5129}

\bibitem{review-de}
 T. Padmanabhan, \textit{Cosmological Constant --- the Weight of the Vacuum},
    Phys. Rept. \textbf{380} (2003) 235--320, \hepth{0212290};\\
 T. Padmanabhan,
{\it Dark Energy: Mystery of the Millennium}, AIP Conf. Proc.
\textbf{861} (2006)  179--196, \astroph{0603114};\\
 P.~Frampton, \textit{Dark Energy --- a Pedagogic Review}, \astroph{0409166};\\
 E.J. Copeland, M. Sami, and  Sh. Tsujikawa,
\textit{Dynamics of dark energy}, Int. J. Mod. Phys. \textbf{D 15}
(2006) 1753--1936, \hepth{0603057};\\
  A.~Albrecht {\it et al.},
  \textit{Report of the Dark Energy Task Force},
  \astroph{0609591};\\
R. Durrer  and R. Maartens, \textit{Dark energy and dark gravity: theory
overview}, Gen. Rel. Grav. \textbf{40} (2008) 301--328,
\arXivid{0711.0077}

\bibitem{ZhangGui} Jingfei Zhang and Yuan-Xing Gui,
\textit{Reconstructing quintom from WMAP 5-year observations:
Generalized ghost condensate}, \arXivid{0910.1200}

\bibitem{Quinmodrev1}
Yi-Fu Cai,  E.N. Saridakis, M.R. Setare, and Jun-Qing Xia,
\textit{Quintom Cosmology: theoretical implications and observations},
\arXivid{0909.2776}\\
 Hongsheng Zhang, \textit{ Crossing the
phantom divide},  \arXivid{0909.3013}

\bibitem{Rubakov}
V.A. Rubakov, \textit{Phantom without UV pathology}, Theor. Math. Phys.
\textbf{149} (2006) 1651--1664 [Teor. Mat. Fiz. \textbf{149} (2006)
409--426],
\hepth{0604153};\\
M. Libanov, E. Papantonopoulos, V. Rubakov, M. Sami, and Sh. Tsujikawa,
\textit{UV stable, Lorentz-violating dark energy with transient phantom
era}, JCAP \textbf{0708} (2007) 010, \arXivid{0704.1848}


\bibitem{Brane}
V. Sahni and Yu. Shtanov, \textit{Braneworld models of dark energy},
JCAP \textbf{0311} (2003) 014,
\astroph{0202346};\\
A. Lue and G.D. Starkmann, \textit{How a brane cosmological constant
can trick us into thinking that $w < -1$},   Phys. Rev. \textbf{D 70}
(2004) 101501,
\astroph{0408246};\\
A.S.~Koshelev and  Th.N.~Tomaras, \textit{Towards a covariant model for
cosmic self-acceleration},
 JHEP \textbf{0710} (2007) 012, \arXivid{0706.3393};\\
M.R. Setare and  E.N. Saridakis, \textit{Braneworld models with a
non-minimally coupled phantom bulk field:
a simple way to obtain the $-1$-crossing at late times},  JCAP \textbf{0903} (2009) 002, \arXivid{0811.4253};\\
Yu. Shtanov, V. Sahni, A. Shafieloo, and  A. Toporensky,
\textit{Induced cosmological constant and other features of asymmetric
brane embedding}, JCAP \textbf{0904} (2009) 023,  \arXivid{0901.3074}

\bibitem{string-cosmo} F. Quevedo, \textit{Lectures on string/brane cosmology},
Class. Quant. Grav. \textbf{19} (2002) 5721--5779, \hepth{0210292};\\
U.H. Danielsson, \textit{Lectures on string theory and cosmology},
Class. Quant. Grav. \textbf{22} (2005) S1-S40,  \hepth{0409274};\\
    M. Trodden and S.M. Carroll, \textit{TASI Lectures: Introduction to Cosmology}, \astroph{0401547};\\
    A. Linde, \textit{Inflation and String Cosmology},
    J. Phys. Conf. Ser. \textbf{24} (2005) 151--160, \hepth{0503195};\\
C.P. Burgess, \textit{Strings, Branes and Cosmology: What can we hope to learn?}, \hepth{0606020};\\
        J.M.~Cline, \textit{String Cosmology},  \hepth{0612129};\\
  L. McAllister and  E. Silverstein, \textit{String Cosmology: A Review},
  Gen. Rel. Grav. \textbf{40} (2008) 565--605,  \arXivid{0710.2951}

\bibitem{0605265}  I.P. Neupane, {\it Towards Inflation and Accelerating
Cosmologies in String-Generated Gravity Models}, \hepth{0605265}

\bibitem{Biswas} T. Biswas, A. Mazumdar, and  W. Siegel,
\textit{Bouncing Universes in String-inspired Gravity}, \textit{JCAP}
\textbf{0603} (2006) 009, \hepth{0508194}


\bibitem{nonlocal}
S. Deser and R.P. Woodard, \textit{Nonlocal Cosmology}, Phys. Rev.
Lett.
\textbf{99} (2007) 111301, \arXivid{0706.2151};\\
S. Nojiri and S.D. Odintsov, \textit{Modified non-local-$F(R)$ gravity
as the key for the inflation and dark energy}, Phys. Lett. \textbf{B
659}
(2008) 821--826, \arXivid{0708.0924};\\
S. Jhingan, S. Nojiri, S.D. Odintsov, M. Sami, I Thongkool, and S.
Zerbini, \textit{Phantom and non-phantom dark energy: The cosmological
relevance of non-locally corrected gravity}, Phys. Lett. \textbf{B 663}
(2008)
424--428, \arXivid{0803.2613};\\
T.S. Koivisto, \textit{Newtonian limit of nonlocal cosmology},  Phys.
Rev. \textbf{D 78} (2008) 123505, \arXivid{0807.3778};\\
S. Capozziello, E. Elizalde, Sh. Nojiri, and S.D. Odintsov,
\textit{Accelerating cosmologies from non-local higher-derivative
gravity}, Phys. Lett. \textbf{B 671} (2009) 193--198,
\arXivid{0809.1535}; \\
F.W. Hehl and B. Mashhoon, \textit{A formal framework for a nonlocal
generalization of Einstein's theory of gravitation}, Phys. Rev.
\textbf{D 79} (2009) 064028, \arXivid{0902.0560};\\
S. Nesseris and A. Mazumdar, \textit{Newton's constant in $f(R,R_{\mu
\nu}R^{\mu \nu}$,$\Box R$) theories of gravity and constraints from
BBN}, Phys. Rev. \textbf{D 79} (2009) 104006, \arXivid{0902.1185};\\
C. Deffayet and R.P. Woodard, \textit{Reconstructing the Distortion
Function for Nonlocal Cosmology}, JCAP \textbf{0908} (2009) 023, \arXivid{0904.0961};\\
 G. Cognola, E. Elizalde, S. Nojiri, S.D. Odintsov, and S. Zerbini,
\textit{One-loop effective action for non-local modified Gauss-Bonnet
gravity in de Sitter space}, Eur. Phys. J \textbf{C 64} (2009)
483--494, \arXivid{0905.0543}

\bibitem{PaisU} A. Pais and  G.E. Uhlenbeck,
\textit{On Field Theories with Nonlocalized Action},
 Phys. Rev. \textbf{79} (1950) 145--165

\bibitem{STinspired}
 D.A. Eliezer and  R.P. Woodard, \textit{The Problem of
Nonlocality in
String Theory}, Nucl. Phys. \textbf{B 325} (1989) 389--469;\\
J. Llosa and J. Vives, \textit{Hamiltonian formalism for nonlocal
Lagrangians},
J. Math. Phys. \textbf{35} (1994) 2856--2877; \\
R.P. Woodard, \textit{A Canonical Formalism For Lagrangians With
Nonlocality Of Finite Extent}, Phys. Rev. \textbf{A 62} (2000) 052105;\\
 K. Bering, \textit{A Note on
Non-Locality and Ostrogradski's
Construction}, \hepth{0007192};\\
 N. Moeller and  B. Zwiebach, \textit{Dynamics with
Infinitely Many Time Derivatives and Rolling Tachyons},  JHEP {\bf
0210} (2002)
034, \hepth{0207107};\\
Ya.I. Volovich, {\it Numerical study of nonlinear
 equations with infinite number of derivatives},
 J. Phys.  {\bf A 36} (2003) 8685--8702, \mathph{0301028};\\
V.S. Vladimirov and  Ya.I. Volovich, \textit{Nonlinear Dynamics
Equation in p-Adic String Theory}, Theor. Math. Phys. {\bf 138} (2004)
297--309 [Teor. Mat. Fiz., {\bf 138} (2004)
355--368], \mathph{0306018};\\
A. Sen, {\it Tachyon Dynamics in Open String Theory},
 Int. J. Mod. Phys. \textbf{A 20} (2005) 5513--5656,
\hepth{0410103}\\
V.S. Vladimirov, {\it On the equation of the $p$-adic open string
for the scalar tachyon field},  \mathph{0507018}; \\
V. Forini, G. Grignani, and  G. Nardelli,
 {\it A new rolling tachyon solution of cubic string field theory},
 JHEP \textbf{0503} (2005) 079, \hepth{0502151};\\
L.V.~Joukovskaya, {\it Iteration method of solving nonlinear integral
equations describing rolling solutions in string theories}, Theor.
Math. Phys. {\bf 146} (2006)  335--342
[Teor. Mat. Fiz. {\bf 146} (2006) 402--409],  \arXivid{0708.0642};\\
B. Dragovich, \textit{Zeta Nonlocal Scalar Fields}, Theor. Math. Phys.
\textbf{157} (2008) 1671--1677
 [Teor. Mat. Fiz. {\bf 157} (2008) 364--372], \arXivid{0804.4114};\\
G. Calcagni and G. Nardelli, \textit{Kinks of open superstring field
theory}, Nucl. Phys. \textbf{B 823} (2009) 234--253,
\arXivid{0904.3744}

\bibitem{Yang} H. Yang,
{\it Stress tensors in p-adic string theory and truncated OSFT}, JHEP
\textbf{0211} (2002) 007, \hepth{0209197}


\bibitem{AJK}  I.Ya. Aref'eva, L.V. Joukovskaya, and  A.S. Koshelev,
\textit{Time evolution in superstring field theory on nonBPS brane. 1.
Rolling tachyon and energy momentum conservation},
 JHEP \textbf{0309} (2003) 012, \hepth{0301137}


\bibitem{davis}
H.T. Davis, \textit{The Laplace differential equation of infinite
order}, Ann. of Math. {\bf 2} 32 (1931) no. 4, 686--714;\\
 H.T. Davis, \textit{The Theory of Linear Operators from the
Standpoint of Differential Equations of Infinite Order}, Indiana, The
Principia Press, 1936

\bibitem{carmi}
R.D. Carmichael, \textit{Linear differential equations of infinite
order}, Bull. Amer. Math. Soc.  \textbf{42} (1936)  193--218;\\
 L. Carleson, \textit{On infinite differential equations with constant
coefficients. I}, Math. Scand.  \textbf{1} (1953) 31--38



\bibitem{ABJV0903}
I.Ya. Aref'eva, N.V. Bulatov, L.V. Joukovskaya, and S.Yu. Vernov,
\textit{Null Energy Condition Violation and Classical Stability in the
Bianchi I Metric},   Phys. Rev. D \textbf{80} (2009) 083532,
\arXivid{0903.5264}

\end{thebibliography}
\end{document}